\documentclass[aps,prd,nofootinbib,twocolumn,groupaddress,preprintnumbers]{revtex4-1}

\usepackage{amssymb}
\usepackage{amsmath}
\usepackage{epsfig}
\usepackage{breakurl}
\usepackage{bm}
\usepackage{color}
\usepackage{tikz}

\makeatletter

\usepackage[utf8]{inputenc}
\usepackage{dsfont}
\usepackage{shuffle}
\usepackage{array}
\usepackage{amssymb,amsmath}
\usepackage[margin=2.2cm]{geometry}
\usepackage{mathtools}
\usepackage{bbold}
\usepackage{color}

\usepackage[
      colorlinks=true,
      linkcolor=darkblue,  
      urlcolor=blue,    
      filecolor=blue,     
      citecolor=red,
      linktocpage=true,
      pdfstartview=FitV,
      bookmarksopen=true    
      ]{hyperref}

\definecolor{darkblue}{rgb}{0.2, 0, 0.8}
\definecolor{darkred}{rgb}{0.9, 0, 0.1}
\usepackage{tikz}
\usetikzlibrary{arrows,shapes}
\usetikzlibrary{trees}
\usetikzlibrary{patterns}
\usetikzlibrary{matrix,arrows} 				% For commutative diagram
\usetikzlibrary{positioning}				% For "above of=" commands
\usetikzlibrary{calc,through}				% For coordinates
\usetikzlibrary{decorations.pathreplacing}  % For curly braces
\usepackage{pgffor}							% For repeating patterns

\usetikzlibrary{decorations.pathmorphing}	% For Feynman Diagrams
\usetikzlibrary{decorations.markings}
\tikzset{
    vector/.style={decorate, decoration={snake}, draw},
	provector/.style={decorate, decoration={snake,amplitude=2.5pt}, draw},
	antivector/.style={decorate, decoration={snake,amplitude=-2.5pt}, draw},
        smallvector/.style={decorate, decoration={snake,amplitude=1.5pt,post length=0.5mm}, draw},
    fermion/.style={draw=black, postaction={decorate},
        decoration={markings,mark=at position .55 with {\arrow[draw=black]{>}}}},
    fermionbar/.style={draw=black, postaction={decorate},
        decoration={markings,mark=at position .55 with {\arrow[draw=black]{<}}}},
    fermionnoarrow/.style={draw=black},
    gluon/.style={decorate, draw=black,
        decoration={coil,amplitude=4pt, segment length=5pt}},
    scalar/.style={dashed,draw=black, postaction={decorate},
        decoration={markings,mark=at position .55 with {\arrow[draw=black]{>}}}},
    scalarbar/.style={dashed,draw=black, postaction={decorate},
        decoration={markings,mark=at position .55 with {\arrow[draw=black]{<}}}},
    scalarnoarrow/.style={dashed,draw=black},
    electron/.style={draw=black, postaction={decorate},
        decoration={markings,mark=at position .55 with {\arrow[draw=black]{>}}}},
    bigvector/.style={decorate, decoration={snake,amplitude=4pt}, draw},
    arrow/.style={draw=black, postaction={decorate},
        decoration={markings,mark=at position 1 with {\arrow[draw=black]{>}}}},
}

\tikzstyle{block} = [draw, rectangle, 
    minimum height=3em, minimum width=6earticlem]

\renewcommand{\r}{\rho}

\newcommand{\reef}[1]{(\ref{#1})}

\def\be{\begin{equation}}
\def\ee{\end{equation}}
\def\bea{\begin{eqnarray}}
\def\eea{\end{eqnarray}}
\def\ba{\begin{array}}
\def\ea{\end{array}}
\def\bd{\begin{displaymath}}
\def\ed{\end{displaymath}}

\def\Tr{{\rm Tr}}

\def\r{\rho}                                     %     \varrho
\def\G{\Gamma}

\def\>{\rangle} %right angle
\def\<{\langle} %left angle
\def\Dsl{D \hskip-.6em \raise1pt\hbox{$ / $ } }
\def\to{\rightarrow}

\newcommand{\bes}{\begin{split}}
\newcommand{\ens}{\end{split}}

\begin{document}

\title{Emergence of String Monodromy in Effective Field Theory}
\author{Alan Shih-Kuan Chen}
\author{Henriette Elvang}
\author{Aidan Herderschee}
\affiliation{Leinweber Center for Theoretical Physics, Randall Laboratory of Physics\\ The University of Michigan, Ann Arbor, MI 48109-1040, USA}
\email{shihkuan@umich.edu,\,elvang@umich.edu,\,aidanh@umich.edu}

\begin{abstract}
String monodromy is a set of linear relations among open string tree amplitudes with different orderings of the vertex operators. In this Letter, we show how these intrinsically stringy relations emerge in low-energy effective field theory from the assumptions of locality and the field theory Kleiss-Kuijf (KK) and Bern-Carrasco-Johansson (BCJ) relations. Specifically, we study the bi-adjoint scalar model effective field theory (BAS EFT). We impose the field theory KK and BCJ relations on one of the two color-orderings of the BAS EFT amplitudes and show that the second color-ordering has an emergent set of linear relations that, as checked at 4-point to 36-derivative order, are exactly the monodromy relations. The 4-point results depend on a delicate interplay between consistent factorization of 6-point BAS EFT amplitudes and the 6-point KK BCJ conditions. As a consequence of the analysis, the 4-point Z-theory tree amplitudes are bootstrapped up to a symmetric function in $s,t,u$ whose simple exponentiated form has free parameters that capture all the odd zeta values when matched to Z-theory.

\end{abstract}

\preprint{LCTP-22-18}

\maketitle

\section{Introduction} 

Type I open string tree-level  scattering processes are  computed as disk amplitudes with $n$ vertex operators inserted on the boundary. Amplitudes with distinct external ordering of the vertex operators correspond to  different choices of contours in the integrals 
over the insertion points. The  linear relations that can be derived from contour deformations are  the {\em string monodromy relations}  \cite{Plahte:1970wy,Bjerrum-Bohr:2009ulz, Stieberger:2009hq,Bjerrum-Bohr:2010mia,Bjerrum-Bohr:2010pnr}. We work here with massless external states for which the 4-point  monodromy relations take the form
\be
  \label{monodromy}
   A_4[1324] + e^{i \pi \alpha' u} A_4[1234] + e^{-i \pi \alpha' t} A_4[1342]  = 0\,.\,
\ee
Exponentials of Mandelstams appear because the integrand is not single-valued on the string moduli space when one takes the necessary contour deformations. Crucially, the monodromy relations are a consequence of the string worldsheet and do not have an obvious origin in the low-energy expansion: for a general low energy EFT, there is no reason for the tree amplitudes to obey monodromy relations. 

At  low-energy (small $\alpha'$-expansion), the string monodromy relations reduce to the field theory  Kleiss-Kuijff (KK) \cite{Kleiss:1988ne} and Bern, Carrasco, Johansson (BCJ) relations \cite{Bern:2008qj}. The KK relations include reversal identities such as $A_n[123\dots n] = (-1)^n A_n[1n \dots 32]$ and the $U(1)$ decoupling identity. 
The BCJ relations are related to color-kinematic duality \cite{Kawai:1985xq,Bern:2008qj}. 
These $(n-1)! -(n-3)!$ conditions, collectively  denoted as ``KKBCJ relations", are field theoretic in nature.
They are known to be obeyed by the tree amplitudes $A_n$ of Yang-Mills theory, super Yang-Mills theory, and the nonlinear sigma-model of chiral perturbation theory.
 
In this Letter, we show that imposing the field theory KKBCJ relations on one color ordering of a $d$-dimensional ($d>3$) massless bi-adjoint scalar effective field theory (BAS EFT), the string monodromy relations automatically emerge at 4-point as linear relations associated with the other color-ordering. This is  unexpected because the monodromy relations are intrinsically stringy. Moreover, the field theory KKBCJ relations leave multiple couplings of higher-derivative operators unfixed in the BAS EFT, yet the emergent linear relations  depend only on a single variable which can be identified as the scale of $\alpha'$. The emergence of the string monodromy relations does not follow from a solely 4-point analysis but relies on the non-trivial factorization of 6-point into 4-point combined with the 6-point KKBCJ constraints.   

\section{Set Up}
\noindent {\bf KKBCJ relations.}
At $n$-point, we have $(n-1)!$ cyclic color-orderings $\alpha$ and there are $(n-1)!$-$(n-3)!$ KKBCJ relations. The linear KKBCJ relations can be written compactly as 
\be
\label{KKBCJvec}
  \sum_{\alpha} A_n[\alpha] \, N^I_n[\alpha] = 0\,,
\ee
in terms of $(n-1)!$-component vectors $N^I_n[\alpha]$ labeled by $I = 1,2,\ldots, (n-1)! -(n-3)!$. At 3-point, there are only two color orderings $\{123, 132\}$ and the only  KKBCJ relation is the reflection identity $N^1_3 = (1,1)$ that says $A_3[123] = -A_3[132]$. Choosing the 6 color-orderings at 4-point to be $\{1234, 1243, 1324, 1342, 1423, 1432\}$, the five KKBCJ vectors (3 reflection identities, the $U(1)$  decoupling relation, and a BCJ relation) can be written as:
\bea
\nonumber
  N_4^1 \,=\, (1, 0, 0, 0, 0, -1)\,,
  && ~N_4^4 \,=\, (1, 1, 1, 0, 0, 0)\,, 
  \\
\nonumber
  N_4^2 \,=\, (0, 1, 0, -1, 0, 0)\,,
  && ~N_4^5 \,=\, (u, -t, 0, 0, 0, 0)\,, \\
  N_4^3 \,=\, (0, 0, 1, 0,-1, 0)\,,
\label{4ptNulls}
\eea
for Mandelstams $s=s_{12}$, $t=s_{13}$, $u=s_{14}$ with $s+t+u = 0$ and $s_{ij} = (p_i+p_j)^2$. The higher-point  KKBCJ vectors  can be found from the expressions in Section 2 of \cite{Bern:2019prr}.

%%%%%%%%%%%%%%%%%%%%%%%%%%%
%%%%%%%%%%%%%%%%%%%%%%%%%%%
%%%%%%%%%%%%%%%%%%%%%%%%%%%
\vspace{2mm}
\noindent {\bf Bi-Adjoint Scalar (BAS) EFT.}
We consider a Lorentz-invariant local $d$-dimensional model with a massless bi-adjoint scalar field $\phi^{aa'}$. It carries adjoint indices under two non-abelian global `color'-symmetry groups, $G$ and $G'$. We assume the model to have a canonical kinetic term and the interactions of interest are single-trace in each color-structure. The tree amplitudes $m_n[\alpha|\beta]$ are doubly color-ordered, with $\alpha$ ($\beta$) denoting the single-trace ordering of $n$ generators associated with $G$ ($G'$). 

We impose the KKBCJ relations on the second color-ordering, i.e.~for each of the $(n-1)!-(n-3)!$ KKBCJ vectors $N^I_n$, we require
\be \label{KKBCJonm}
  \sum_{\beta} m_n[\alpha|\beta] \, N^I_n[\beta] = 0\,,
\ee
where the sum running over the $(n-1)!$ cyclically independent color-orderings $\beta$.  

The KKBCJ condition 
 in Eq.~\reef{KKBCJonm} is the statement that the $(n-1)! \times (n-1)!$ matrix $\mathbf{m}_n$ of tree amplitudes $m_n[\alpha|\beta]$ has $(n-1)!-(n-3)!$ null vectors $N^I_n$ under right-multiplication. It follows that $\mathbf{m}_n$ must have rank $(n-3)!$ and therefore  $\mathbf{m}_n$ must also have  $(n-1)!-(n-3)!$ null vectors $K^I_n$ under left-multiplication, 
\be \label{Knulls}
  \sum_{\alpha} K^I_n[\alpha] \, m_n[\alpha|\beta]   = 0\,.
\ee
We show in this Letter that when the BAS EFT model is allowed to have its most general local interactions, the null vectors $K^I_n$ at 4-point turn out to be exactly the string monodromy relations. 

\section{Solving KKBCJ}
The KKBCJ constraints in Eq.~\reef{KKBCJonm} are straightforward to solve order by order in the derivative expansion. We now describe the analysis at 3-, 4-, 5-, and 6-points. 

\vspace{1mm} 
\noindent {\bf 3-point.}
The bi-adjoint scalar can have two possible cubic self-interactions: one in which the 3 pairs of adjoint indices are contracted with the anti-symmetric structure constants $f^{abc}$ and another with the fully symmetric tensors $d^{abc}  =\Tr[T^a \{T^b,T^c\}]$. The reversal symmetry constraint requires $m_3[123|123] = - m_3[123|132]$ which rules out the 
 $d^{abc}$-tensor cubic interaction. In the absence of other interactions, the result is the cubic bi-adjoint scalar model
 \be
   \mathcal{L}_\text{BAS} = -\frac{1}{2} (\partial \phi^{aa'})^2 + \frac{1}{6}g f^{abc} \tilde{f}^{a'b'c'} \phi^{aa'} \phi^{bb'} \phi^{cc'} \,.
\ee
The cubic BAS model satisfies the KKBCJ relations for both color-orderings. 
 
 \vspace{1mm} 
We now consider all possible local single-trace (in $G$ and $G'$) higher-derivative interactions, schematically $\text{tr}\,\partial^{2k}\phi^n$. On-shell, the $n$-field Lagrangian operators of this form are one-to-one with polynomial terms in the $n$-point matrix elements. 
 
\vspace{1mm} 
\noindent {\bf 4-point.} Using cyclicity and momentum relabeling, the $3! \times 3! = 36$ possible color-ordered amplitudes  $m_4[\alpha|\beta]$ can be written 
 in terms of three functions \cite{Chi:2021mio}:  
\bea
  \nonumber
  &&\!\!\!f_1(s,t) = m_4[1234|1234]\,,  ~ f_2(s,t) = m_4[1234|1243]\,,
  \\  
  &&\!\!\!f_6(s,t) = m_4[1234|1432]\,,
\eea
We impose the KKBCJ relations from Eqs.~\reef{4ptNulls}-\reef{KKBCJonm}. $N_4^1$  gives $f_6(s,t) = f_1(s,t)$ and then $N_4^2$ and $N_4^3$ are automatically satisfied. The $N_4^4$ and $N_4^5$ constraints are solved by
\be
\label{f2cond}
f_1(s,t) = \frac{t}{u} f_2(s,t)
~~~\text{and}~~~
u f_2(u, s) = t f_2(t, s) \,.
\ee
These relations ensure that $f_1$ is cyclic, $f_1(u,t)=f_1(s,t)$. Therefore we write the most general ansatz for  $f_2$ and impose Eq.~\reef{f2cond}. Setting the cubic coupling $g=1$ without loss of generality, we write 
\be 
  \label{f2ansatz}
   f_2(s,t) = -\frac{1}{s} 
   + \sum_{k=0}^N \sum_{r=0}^k
   a_{k,r} \,s^r \,t^{k-r} \ ,
\ee
where terms up to $2N$ derivatives are included. Solving the second condition of Eq.~\reef{f2cond} order-by-order in the momentum expansion, we find
{\small
\bea
\nonumber
a_{0,0}= 0, \hspace{2.75cm}
&&
~a_{1,1} = a_{1,0}, 
\\ \nonumber
a_{2,1} = a_{2,0}, \hspace{2.35cm}
&& 
~a_{2,2} = 0, 
\\
\label{4ptsol}
a_{3,2} =  2a_{3,1}-2a_{3,0}, \hspace{1.1cm}
&&
~a_{3, 3}=a_{3, 1}-a_{3, 0},
\\ \nonumber
a_{4, 2} =  2 a_{4, 1}-2a_{4, 0},\hspace{1.1cm}
&&
~a_{4, 3} =a_{4, 1}-a_{4, 0},
\\ \nonumber
a_{4, 4} =0, \hspace{2.7cm}
&&
~a_{5,3} = 5a_{5, 0} - 5 a_{5,1} + 3 a_{5,2},
\\ \nonumber
a_{5,4} =6 a_{5,0} - 6 a_{5,1} + 3 a_{5,2},
&&
~a_{5,5} =2 a_{5,0} - 2 a_{5,1} + a_{5,2},
\eea}%
and it is straightforward to extend to higher order. There is no mixing between orders since the KKBCJ relations are linear equations. Moreover, $f_1$ as given by Eq.~\reef{f2cond} does not have unphysical poles. 

As argued above, the fact that $\mathbf{m}_4$ has rank 1 means that there are 5 left-multiplication null vectors $K_4^I$ of Eq.~\reef{Knulls}. They consist of the 3 reversal-symmetry null vectors $K_4^I = N_4^I$, $I=1,2,3$, and two other null vectors that are  generalizations of the $U(1)$ decoupling identity and of the BCJ relations. By Ref.~\cite{Chi:2021mio} 
\be 
 \label{K4K5}
 \begin{split}
  K_4^4 &= \Big\{1, 1,  -\frac{f_2(s, u)}{f_2(u, s)} - \frac{f_2(s, t)}{f_2(t, s)},
   0, 0, 0\Big\}\,,\\
  K_4^5 &=   \Big\{ 1, -\frac{f_2(u, t)}{f_2(t, u)}, 0, 0, 0, 0 \Big\} \,.
  \end{split}
\ee
The 4-point KKBCJ relations fix about half of the $a_{k,r}$'s, and in general the null vectors in Eq.~\reef{K4K5} depend on all of the free $a_{k,r}$'s in $f_2$.  
This changes when constraints from the 6-point analysis are implemented. 

\begin{figure*}[t!]
  \begin{eqnarray}
  &&
  \label{BCJrel1}
  \begin{split}
0 &= s_{34}s_{234} \, m_6[\mathbb{1}|134265] + s_{46}(s_{12}+s_{25}) \, m_6[\mathbb{1}|136425] + s_{46} s_{12} \, m_6[\mathbb{1}|136452] + (s_{46}+s_{45})s_{12} \, m_6[\mathbb{1}|136542] \\
&~~~+ (s_{14}+s_{45})(s_{234}+s_{26}) \, m_6[\mathbb{1}|136245] + s_{14}(s_{234}+s_{26}) \, m_6[\mathbb{1}|136254] 
+ s_{14}(s_{234}+s_{26}+s_{25}) \, m_6[\mathbb{1}|136524] \,~~~\phantom{1}
\end{split}\\[1mm]
\implies
&&
 \label{BCJrel1SK}
  0 = \Big[ s_{34}\, f_2(s_{34},s_{24})%[34P?2342] 
 \,
 f_2(s_{56},s_{15}) %[1P23456] 
 - s_{45}  \, g_4[P_{12} 3456]   \Big] \Big|_{s_{14}, s_{26}, s_{46},s_{234}, s_{12} = 0}\,,
 %{\overset{\scriptstyle{s_{14}, s_{64}, s_{62},}}{s_{234}, s_{12}} = 0}
 \\[2mm]
 &&
\label{BCJrel2}
\begin{split}
0 &= s_{23}s_{234} \, m_6[\mathbb{1}|132456] + s_{25}(s_{14}+s_{46}) \, m_6[\mathbb{1}|135246] 
+ s_{25} s_{14} \, m_6[\mathbb{1}|135264] 
+ (s_{25}+s_{26})s_{14} \, m_6[\mathbb{1}|135624]\\
&~~~~+ (s_{12}+s_{26})(s_{234}+s_{45}) \, m_6[\mathbb{1}|135426] 
+ s_{12}(s_{234}+s_{45}) \, m_6[\mathbb{1}|135462] 
+ s_{12}(s_{234}+s_{45}+s_{46}) \, m_6[\mathbb{1}|135642]~~~~\phantom{~}
\end{split}
\\
\implies
&&
 \label{BCJrel2SK}
 0 = \Big[ s_{23} \, \big( f_{2}(s_{23},s_{24}) \, f_{1}(s_{16},s_{15}) \big) 
 - s_{26} \, g_{3}[6123P_{45}] 
 +  f_{1}(s_{123},s_{36})- s_{46}  \, g_{3}[P_{12}3456]  \Big] \Big|_{\overset{\scriptstyle{s_{25}, s_{14}, s_{234},}}{s_{12}, s_{45}} = 0}~~
  \end{eqnarray} 
\caption{\label{fig:2BCJs} The BCJ relations in Eq.~(4.27) of Ref.~\cite{Bern:2008qj} are imposed on the $\beta$ color-ordering of the BAS EFT tree amplitudes $m_6[\alpha|\beta]$ for any choice of the color-ordering $\alpha$. Eq.~\reef{BCJrel1} is obtained by choosing $\alpha = 154263$ and then relabeling $[154263] \to [123456]$ and Eq.~\reef{BCJrel2} arises from $\alpha = 142536$ and then relabeling $[142536]\to [123456]$. The notation $\mathbb{1}$ is shorthand for the color-ordering $123456$. 
To get Eq.~\reef{BCJrel1SK}, take the limit $s_{14}$, $s_{26}$, $s_{46} \to 0$ of Eq.~\reef{BCJrel1} (the amplitudes have no poles in those variables) and then pickup the residues from poles at  $s_{234} =0$ and $s_{12} = 0$. The identity in Eq.~\reef{BCJrel1SK} precisely realizes the relation between diagrams shown in Eq.~\reef{sample6pt}. 
A similar kinematic limit of Eq.~\reef{BCJrel2} gives Eq.~\reef{BCJrel2SK}.} 
\end{figure*}

\vspace{1mm} 
\noindent {\bf 5-point.} The $4! \times 4!$ amplitudes of $\mathbf{m}_5$ can be parameterized in terms of 8 basis amplitudes $g_1,\ldots, g_8$ defined in Eq.~(6.1) of Ref.~\cite{Chi:2021mio}. For each $g_i$, we construct a consistent factorization to 3- and 4-point and then include all possible polynomial terms in the Mandelstams to parameterize the possible local contact terms. Their coefficients are constrained by imposing the 22 5-point KKBCJ relations in Eq.~\reef{KKBCJonm}, which we have solved to $O(s^{17})$. As an example, relevant in the following,  the result for $g_4 \equiv m_5[12345|12543]$ is
 {\footnotesize
 \bea
 \nonumber
 \!\!&&g_4[12345] 
   \frac{1}{s_{12} s_{34}}
 + \frac{1}{s_{12} s_{45}}
 + a_{1,0}\Big(
 - \frac{s_{35}}{s_{12}}
 +  \frac{s_{15}}{s_{34}}  + \frac{s_{23}}{s_{45}} \Big)\\
 \!\!&&
 ~~+ a_{2, 0} \Big( 
   - \frac{s_{35}^2}{s_{12}} 
   + \frac{s_{15} s_{25}}{s_{34}} 
   + \frac{s_{13} s_{23}}{s_{45}} 
   + 2 s_{35}  \Big) + \ldots .
   \label{g4res}
 \eea}%
At $O(s^2)$, the expressions are longer, but all terms in the $g_i$ are fixed at that order in terms of $a_{3,0}$  and $a_{3,1}$. At $O(s^3)$ all local coefficients, except two, are fixed in terms of $a_{4,0}$ and  $a_{4,1}$. At $O(s^4)$ there are five free parameters, one of which parameterizes a local term that violates reversal symmetry  $m_n[\alpha^T|\beta] = (-1)^n m_n[\alpha|\beta]$, where $\alpha^T$ is the reverse of the color-order $\alpha$ (e.g.~$(12345)^T = (54321)$). Importantly, we find no constraints on the $a_{k,r}$ from the 5-point KKBCJ relations.

\vspace{1mm} 
\noindent {\bf 6-point.}  The $5! \times 5!$ amplitudes of the matrix $\mathbf{m}_6$ are written in terms of 24 basis amplitudes that are independent under cyclicity and momentum relabeling. We construct the most general ansatz for the basis amplitudes consistent with all factorization channels and with arbitrary coefficients for the local 6-point contact terms. The ansatz is subjected to the 114 6-point KKBCJ relations and solved systematically order-by-order. Remarkably one  finds that the 4-point coefficients $a_{k,r}$ are constrained; for example, we find that 
\be
  \label{a3res}
  a_{3,0} = -\frac{2}{5} a_{1,0}^2 \,,~~~
  a_{3,1} = -\frac{11}{10} a_{1,0}^2 \,.
\ee
Together with the 4-point constraints in Eq.~\reef{4ptsol} this completely fixes the $O(s^3)$ terms in the $f_2$ ansatz of Eq.~\reef{f2ansatz} in terms of $a_{1,0}$.

Since the KKBCJ relations are linear,  non-linear relations, such as in Eq.~\reef{a3res}, that mix orders in the 4-point expansion may seem surprising. It is clear that they have to arise from the factorization channels in the 6-point amplitudes, but diagrams with the same pole-structure cannot possibly mix orders. So the non-linear relations have to come from a combination of diagrams with different pole structure. With their explicit Mandelstam factors, the BCJ relations allow such mixing of diagrams, so this points towards looking at kinematic limits of the BCJ diagrams that isolate pole-contributions from diagrams  such as
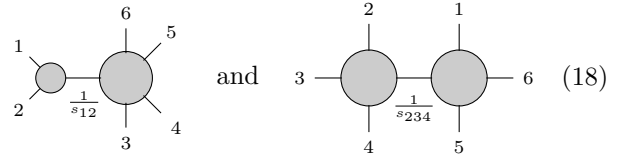
\begin{equation}\label{sample6pt}
\begin{split}
\pgfmathsetmacro{\r}{0.8}
\begin{tikzpicture}[baseline={([yshift=-.5ex]current bounding box.center)},every node/.style={font=\scriptsize}]
\def\dis{1};
\def\rad{0.33};
\def\radb{0.2};
\def\mul{2};
\def\rb{0.35};
\def\sh{0.2};
\node at (0,0) (bub1) [draw, fill=black!20, circle, inner sep=4] {\phantom{$g_{4}$}}; 
\draw (0,\rad) -- (0,\mul*\rad);
\node at (0,\mul*\rad+\sh) {$6$};
\draw (0,-\rad) -- (0,-\mul*\rad);
\node at (0,-\mul*\rad-\sh) {$3$};
\draw (0.707*\rad,-0.707*\rad) -- (\mul*0.707*\rad,-\mul*0.707*\rad);
\node at (\mul*0.707*\rad+\sh,-\mul*0.707*\rad-\sh) {$4$};
\draw (0.707*\rad,0.707*\rad) -- (\mul*0.707*\rad,\mul*0.707*\rad);
\node at (\mul*0.707*\rad+0.7*\sh,\mul*0.707*\rad+0.7*\sh) {$5$};
\node at (-\dis,0) (bub1) [draw, fill=black!20, circle, inner sep=4] {$$};
\draw (-0.707*\radb-\dis,0.707*\radb) -- (-\mul*0.707*\radb-\dis,\mul*0.707*\radb);
\node at (-\mul*0.707*\radb-\dis-0.7*\sh,\mul*0.707*\radb+0.7*\sh) {$1$};
\draw (-0.707*\radb-\dis,-0.707*\radb) -- (-\mul*0.707*\radb-\dis,-\mul*0.707*\radb);
\node at (-\mul*0.707*\radb-\dis-0.7*\sh,-\mul*0.707*\radb-0.7*\sh) {$2$};
\draw (\radb-\dis,0) -- (-\rad,0);
\node at (-\dis/1.8,-\rb) {$\frac{1}{s_{12}}$};
\end{tikzpicture} 
\quad \text{and} \quad 
\begin{tikzpicture}[baseline={([yshift=-.5ex]current bounding box.center)},every node/.style={font=\scriptsize}]
\def\dis{1.2};
\def\rad{0.36};
\def\mul{2};
\def\rb{0.4};
\def\sh{0.2};
\node at (0,0) (bub1) [draw, fill=black!20, circle, inner sep=4] {\phantom{$f_{2}$}};
\draw (0,\rad) -- (0,\mul*\rad);
\node at (0,\mul*\rad+\sh) {$1$};
\draw (0,-\rad) -- (0,-\mul*\rad);
\node at (0,-\mul*\rad-\sh) {$5$};
\draw (\rad,0) -- (\mul*\rad,0);
\node at (\mul*\rad+\sh,0) {$6$};
\node at (-\dis,0) (bub1) [draw, fill=black!20, circle, inner sep=4] {\phantom{$f_{2}$}};
\draw (-\dis,\rad) -- (-\dis,\mul*\rad);
\node at (-\dis,\mul*\rad+\sh) {$2$};
\draw (-\dis,-\rad) -- (-\dis,-\mul*\rad);
\node at (-\dis,-\mul*\rad-\sh) {$4$};
\draw (-\rad-\dis,0) -- (-\mul*\rad-\dis,0);
\node at (-\mul*\rad-\dis-\sh,0) {$3$};
\draw (\rad-\dis,0) -- (-\rad,0);
\node at (-\dis/2,-\rb) {$\frac{1}{s_{234}}$};
\end{tikzpicture}
\end{split}
\end{equation}
Consider the $O(s)$ terms from these diagrams. 
In the first diagram, this comes from the  $O(s^2)$-terms in the 5-point amplitude which are determined completely by $a_{3,0}$ and $a_{3,1}$, as noted below  Eq.~\reef{g4res}. In contrast, the product of 4-point amplitudes in the second diagram has
 contributions with coefficients $a_{1,0}^2$ as well as $a_{3,0}$ and $a_{3,1}$. Thus, a BCJ relation involving two such diagrams can result in a relation such as 
Eq.~\reef{a3res}. 

We can make this argument  precise by isolating the pole contributions in the BCJ relations via  carefully chosen kinematic limits. Figure 1 shows how to do this for two BCJ relations that suffice to derive the needed 4-point constraints.
Expanding Eqs.~\reef{BCJrel1SK} and \reef{BCJrel2SK} in Mandelstams, we find the first non-vanishing terms to be 
\be
 \begin{split}
 0\,=\,& (a_{1,0}^2 - 3 a_{3,0} + 2 a_{3,1}) s_{16} s_{23} s_{34} \,,\\
 0\,=\,& (-a_{1,0}^2 - 8 a_{3,0} + 2 a_{3,1}) s_{23} s_{34} (s_{16} + s_{56}) \,,
 \end{split}
\ee
from which Eq.~\reef{a3res} follows. For the next two powers in the Mandelstams we find
\bea
\nonumber
a_{4,0}=\text{free},\hspace{2.05cm}
  && 
  ~a_{5,1} = \tfrac{34}{35} a_{1, 0}^3   - \tfrac{1}{2}  a_{2, 0}^2,
  \\[1mm]
\nonumber
  a_{4,1} =-a_{1, 0} a_{2, 0} + 2 a_{4, 0},
  &&
  ~a_{5,2}=\tfrac{27}{14} a_{1, 0}^3   -   a_{2, 0}^2,
  \\[1mm]
  \label{6ptsol}
  a_{5,0}= \tfrac{8}{35} a_{1, 0}^3.\hspace{1.65cm}
\eea
Together with the 4-point results, e.g.~\reef{4ptsol}, we have found up to $O(s^{18})$ that Eqs.~\reef{BCJrel1SK} and \reef{BCJrel2SK} fix all 4-point coefficients except $a_{1,0}$ and $a_{2k,0}$. Moreover, some 5-point local coefficients are fixed too by the 6-point KKBCJ constraints; we find in particular that operators that violate reversal symmetry are eliminated.

\section{Emergent Monodromy}

The 4-point string monodromy relation Eq.~\reef{monodromy} can be written in terms of null vectors  $K_4^{I,\text{str}}$ found by setting $f_2 $ in \reef{K4K5} to be $f_2^\text{str}(s,t) = -1/\sin(\pi \alpha'  s)$; see Refs.~\cite{Mizera:2016jhj,Chi:2021mio}.  
For example, using reversal symmetry, $K_4^{5,\text{str}}$ is exactly the imaginary part of Eq.~\reef{monodromy} assuming the amplitudes are real.

Expanding $K_4^{I,\text{str}}$ at low-energies, $\alpha' s_{ij} \ll 1$, we can compare to our left-multiplication null-vectors  $K_4^I$ with the most general $f_2$ found from the 4- and 6-point KKBCJ relations. As a consequence of the 6-point analysis, the only free parameters in the 4-point amplitude $f_2$ are $a_{1,0}$ and $a_{2k,0}$ for $k=1,2,3,\ldots$. We find that  all $a_{2k,0}$  drop out from the ratios of $f_2$ in the left-multiplication null vectors $K_4^I$ in Eq.~\reef{K4K5} which therefore only depend on $a_{1,0}$.  Our $K_4^I$'s exactly match the 4-point string monodromy null-vectors $K_4^{I,\text{str}}$ when 
\be
\label{a10val}
a_{1,0} = - \frac{\pi^2}{6} \alpha'^2 \,,
\ee
as checked up to 18th order in the Mandelstams. Thus, {\em the 4-point string monodromy relations are emergent in the BAS EFT model!} Moreover, we find that the left-multiplication null vectors at 5- and 6-point are likewise the string monodromy relations as checked to $O(s^6)$ and $O(s^4)$, respectively.

\section{Factorization, Exponentiation, Resummation}

We can understand why the dependence on the  $a_{2k,0}$'s drop out from the $K_4^I$'s by examining the structure of $f_2$ more closely. 
It turns out the $f_2$ factorizes: 
\be 
  \label{f2U}
   f_2 = f_2^{(0)} \,U \,,~~~~f_2^{(0)} = f_2\Big|_{a_{2k,0}=0}\,,
\ee
and all the $a_{2k,0}$-dependence is contained in the function $U$ that is fully symmetric in $s,t,u$. Hence ratios of the $f_2$'s in the $K_4^I$ null vectors of Eq.~\reef{K4K5} are independent of $U$, and hence of the $a_{2k,0}$'s.    

Moreover, from the low-energy expansion (to 18th order in the Mandelstams), we find that $U$ can be re-summed into the form $U=e^V$ with
\be
  \label{V}
  V(s,t,u) = \sum_{k=1}^\infty \frac{a_{2k,0}}{2k+1}  \Big(s^{2k+1}+t^{2k+1}+u^{2k+1}\Big)\,.
\ee
Since there can be more than one symmetric polynomial in $s,t,u$ starting at order 6, it is highly non-trivial that $U$ takes this form and it relies heavily on the constraints from the 6-point analysis.

\vspace{2mm}
\noindent {\bf Z-theory.}  
Because the BAS EFT tree amplitudes obey KKBCJ relations on the 2nd color ordering, its amplitudes can be left-sector input in double-copy relations with the field theory KLT kernel Ref.~\cite{Chi:2021mio}. This way, it can for example be double-copied with YM theory to give YM + higher derivative (h.d.) amplitudes:
\be
  \label{YMhd}
  A^\text{YM+h.d.}_n[\alpha] = \sum_{\beta,\gamma} m_n[\alpha|\beta] \,S_n^\text{FT}[\beta|\gamma] \,A^\text{YM}_n[\gamma]\,,
\ee
where  the sum is over a choice two set of $(n-3)!$ color-orderings $\beta$ and $\gamma$, and $S_n^\text{FT}$ denotes the standard field theory KLT kernel. 
The property that $m_4[\alpha|\beta]$ obeys the string monodromy relations on the first color-ordering $\alpha$, is inherited by the YM+h.d.~tree amplitudes constructed by Eq.~\reef{YMhd}. Such amplitudes will include the open string tree amplitudes. 

Indeed, the open string tree amplitudes are known to be constructible by a double-copy such as in Eq.~\reef{YMhd} but with $m_n[\alpha|\beta]$ replaced by the Z-theory amplitudes $m^\text{Z}_n[\alpha|\beta]$. The Z-theory amplitude $f_1$ is the beta-function and $f_2$, found from the BCJ relation, is 
\be
 f_2^\text{Z}(s,t)  =-\frac{1}{s} 
 \frac{\Gamma(1 + \alpha' s) \Gamma(1 - \alpha' (s+t))}{
\Gamma(1 - \alpha' t )}\,.
\ee
The Z-theory amplitudes arise from period integrals in the computation of the disk amplitudes, see Ref.~\cite{Broedel:2013tta}, and they are known to obey string monodromy on the first color-ordering and field theory KKBCJ relations on the second one.

Our BAS EFT model is a generalization of Z-theory; our model  is the most general BAS EFT model that obeys the KKBCJ relations on one color-ordering and we have provided evidence that this is sufficient to imply  string monodromy on the other color-ordering.

This is the special case of our BAS EFT with  $a_{1,0}$ identified with $\alpha'$ as in \reef{a10val} and 
\be
\label{Zakrs}
a_{2k,0} = - (\alpha')^{2k+1} \zeta(2k+1)   \,.
\ee
The entire dependence on the odd-$\zeta$'s is contained in the symmetric function $U$. Note that the exponentiated form \reef{V} with the coefficients of Eq.~\reef{Zakrs} can also be found in Ref.~\cite{Schlotterer:2012ny}. 
For the choice in 
 Eq.~\reef{Zakrs}, the symmetric function $V$ in Eq.~\reef{V} re-sums to a  compact expression with  logarithms of $\G$-functions, and knowing from Eq.~\reef{f2U} that $f_2^Z = f_2^{(0)} e^V$ this leads us to propose a re-summed form of $f_2^{(0)}$, namely
\be 
   \label{f2resummed}
   f_2^{(0)} = - \frac{1}{s}\sqrt{\frac{\pi\, s (s+t) \alpha' \sin(\pi \alpha' t) }{ t\, \sin(\pi \alpha' s) \sin(\pi \alpha' (s+t))}} \, .
\ee 
One can directly verify that gives amplitudes that solve the string monodromy relations on the first color-ordering. 

\section{Discussion.} To derive the monodromy relations, we assumed the cubic coupling and $a_{1,0}$ to be non-zero in the BAS EFT model. If we instead set $a_{1,0} = 0$, the left-multiplication null vectors become the field theory KKBCJ relations instead of the string monodromy relations, and it follows from Eq.~\reef{f2U} that the only solution for $f_2$ is then $-1/s$ times the symmetric function $U$. Choosing $a_{2k,0}$ to be twice the value for Z-theory given in Eq.~\reef{Zakrs} these amplitudes are then the ``$J$-integrals" of the closed string in Ref.~\cite{Stieberger:2014hba}. It is not surprising the $J$-integrals involve $U^2$ of Z-theory because they can be obtained from the double-copy of Z-theory with itself using the string kernel. 

Thanks to linearity,  the sum of any two $n$-point amplitudes that solve $n$-point KKBCJ relations is a again a solution to the KKBCJ relations, however, this may not be compatible with factorization. For example, the sum of two 4-point Z-theory amplitudes with different choices of $\alpha'$ necessarily solve the 4-point KKBCJ relations on the second color-ordering. One can write an ansatz for a 6-point amplitude which correctly factorizes into this 4-point amplitude, but crucially our analysis shows such a 6-point amplitude itself would not be compatible with 6-point KKBCJ relations. This is an interesting interplay between the linear KKBCJ constraints on the amplitudes and the non-linearities introduced by factorization.

We have provided evidence that the string monodromy relations emerge from imposing the KKBCJ relations on the local massless scalar bi-adjoint EFT. These results are strictly tree level, so an obvious question is if there are generalizations at loop level. The proposed monodromy relations for loop integrands studied
in Refs.~\cite{Bjerrum-Bohr:2011jrh,Mafra:2012kh,Tourkine:2016bak,Ochirov:2017jby,Hohenegger:2017kqy,Tourkine:2019ukp,Casali:2019ihm,Casali:2020knc} 
may shed light on this.

An important consequence of the string monodromy relations is that they guarantee that the KLT double-copy of two open string amplitudes is independent of the choice of $(n-3)!$ color-orderings in the double-copy sum; see Refs.~\cite{Johnson:2020pny} and \cite{Chi:2021mio}. The results presented in this Letter have  consequences for generalizations of the double-copy, as will be described in forthcoming work Ref.~\cite{ours2appear}.

\section*{Acknowledgements.}
 We would like to thank Nima Arkani-Hamed, Justin Berman, Lance Dixon, and Sebastian Mizera for useful comments and discussions. AC were supported in part by a Leinweber Summer Fellowship. AC and HE was supported in part by DE-SC0007859.  AH was supported by a Rackham Predoctoral Fellowship from the University of Michigan.

%%%%%%%%%%%%%%%%%%%%

\end{document}